\documentclass[runningheads]{llncs}
\usepackage{graphicx}
\usepackage{hyperref}
\usepackage{booktabs}
\usepackage{longtable}
\usepackage{enumitem}
\usepackage{array}
\usepackage{makecell}
\usepackage{multirow}
\usepackage{url}

\title{From Content Creation to Citation Inflation: A GenAI Case Study}
\titlerunning{AI-generated papers}

\author{
Haitham S.\ Al-Sinani\inst{1}\thanks{\href{https://orcid.org/0009-0005-0453-3335}{ORCID: 0009-0005-0453-3335}} \and
Chris J.\ Mitchell\inst{2}\thanks{\href{https://orcid.org/0000-0002-6118-0055}{ORCID: 0000-0002-6118-0055}}
}
\authorrunning{H. Al-Sinani \and C. Mitchell}

\institute{
\textsuperscript{1}Department of Cybersecurity and Quality Assurance, Diwan of Royal Court, Muscat,  Oman, 
\email{hsssinani@diwan.gov.om} \\
\textsuperscript{2}Department of Information Security, Royal Holloway, University of London, Egham, UK, 
\email{C.Mitchell@rhul.ac.uk}
}

\begin{document}
\maketitle

\begin{abstract}
This paper investigates the presence and impact of questionable, AI-generated academic papers on widely used preprint repositories, with a focus on their role in citation manipulation. Motivated by suspicious patterns observed in publications related to our ongoing research on GenAI-enhanced cybersecurity, we identify clusters of questionable papers and profiles. These papers frequently exhibit minimal technical content, repetitive structure, unverifiable authorship, and  mutually reinforcing citation patterns among a recurring set of authors.  
To assess the feasibility and implications of such practices, we conduct a controlled experiment: generating a fake paper using GenAI, embedding citations to suspected questionable publications, and uploading  it to one such repository (ResearchGate). Our findings demonstrate that such papers can bypass platform checks, remain publicly accessible, and contribute to inflating citation metrics like the H-index and i10-index. We present a detailed analysis of the mechanisms involved, highlight systemic weaknesses in content moderation, and offer recommendations for improving platform accountability and preserving academic integrity in the age of GenAI.
\end{abstract}

\section{Introduction}
Online academic platforms such as ResearchGate  have transformed scholarly communication by enabling researchers to freely share publications, track impact, and engage with a global audience. Their open-access nature and user-driven upload features have democratised knowledge dissemination and expanded the visibility of scientific output. However, this openness has also introduced vulnerabilities. Our investigations suggest that large quantities of AI-generated content are being uploaded to widely used preprint repositories, possibly   with the goal of artificially inflating key metrics for academics, including citation counts, the H-index and the i10-index. 

Citation-based metrics such as the \textit{H-index} and \textit{i10-index} are commonly used to assess a researcher’s academic impact. The H-index is defined as the maximum value \textit{h} such that the researcher has published \textit{h} papers each of which has been cited at least \textit{h} times. It attempts to capture both productivity and citation impact in a single number. The i10-index, introduced by Google Scholar, is a simpler metric that counts the number of a researcher’s publications with at least ten citations. While these metrics are convenient and widely adopted in academic evaluations, they are also vulnerable to manipulation—especially in open publishing environments where content vetting is minimal.

This emerging malpractice is especially troubling in the current technological landscape, where  GenAI  tools ---such as ChatGPT\footnote{https://chatgpt.com/}, Gemini\footnote{https://gemini.google.com/app} and DeepSeek\footnote{https://chat.deepseek.com/},  can effortlessly produce texts that mimic academic style, structure, and tone. With minimal oversight, these synthetic documents can appear convincing enough to be mistaken for legitimate scholarship, especially on platforms where submissions are not peer-reviewed or institutionally validated.

Our investigation is motivated by a fundamental question: \textit{Are there fully AI-generated  papers on ResearchGate, particularly in the area of GenAI-enhanced Cybersecurity, and if so, how are they being used to manipulate academic metrics?} Unlike earlier studies~\cite{2025_CitationManipulationThroughCitationMillsAndPreprintServers_HazemIbrahim,2023_ArtificialIntelligenceCanGenerateFraudulentButAuthenticLookingScientificMedicalArticlesPandoraBoxHasBeenOpened_Medical} that focused on peer-reviewed journals or high-profile conferences, we focus on ResearchGate—a popular but lightly moderated platform where researchers frequently upload preprints, reports, and presentation materials. The consequences of citation manipulation here may be less visible than in traditional publishing, but they are no less significant—particularly because ResearchGate documents are assigned DOIs (Digital Object Identifier) and often indexed by systems like Google Scholar.

Our interest in this topic arose through our own line of research on applying GenAI in the field of ethical hacking. While monitoring related publications via Google Scholar alerts, we encountered a series of papers that appeared, on the surface, to be relevant. However, closer inspection revealed they were strikingly similar in content, cited many of the same individuals, and lacked empirical foundations. These observations raised suspicions that the papers were generated using GenAI and published with the intent of artificially inflating citation metrics.

Motivated by these observations, we conducted a multi-phase investigation to explore this phenomenon more systematically. We began by mapping out and reviewing a cluster of suspicious papers and the ResearchGate profiles associated with them. Building on these findings, we  designed and carried out a controlled experiment: generating a fake academic paper using GenAI, embedding citations to several known suspect papers, and uploading it to ResearchGate under fictional authorship. By observing whether our paper is accepted by the platform and whether it contributes to boosting citation metrics of the referenced papers, we assess the ease with which such manipulation can occur. Our final analysis includes a comparison of multiple ResearchGate profiles suspected of participating in citation inflation schemes.
 
 The goal of this work is not only to document the mechanisms of citation manipulation, but also to raise awareness and prompt a broader discussion on academic integrity in the GenAI age. We present a detailed case study of how generative tools and platform design can interact in ways that compromise the credibility of scholarly metrics. In doing so, we contribute to the growing body of work examining the risks of AI-generated content in academia, and we offer preliminary recommendations for mitigating these threats on platforms like ResearchGate.

This work is timely in that it addresses a growing and under-examined threat to research evaluation systems—namely, the deliberate misuse of GenAI and academic platforms to fabricate publications and manipulate citation metrics. By combining empirical observation with an experimental case study, we provide concrete evidence of how such manipulation can be carried out on ResearchGate. Our findings not only expose systemic gaps in platform oversight but also underscore the need for greater scrutiny of citation-based indicators that are often used in hiring, promotion, and funding decisions.

The remainder of this paper is organised as follows. Section~\ref{ThePhenomenonofFakePapers} introduces the phenomenon of questionable papers, and, section~\ref{PossibleMotivationsforFakePaperGeneration} outlines the core motivations behind their generation. Section~\ref{Preliminary Experiment} presents our experimental methodology, including the design, generation, and upload of a fake academic paper. Section~\ref{ImplicationsAndRecommendations} discusses the broader implications of our findings and provides policy-level recommendations. Section~\ref{RelatedWork} reviews relevant prior studies.  Section~\ref{ConclusionandFutureWork} concludes the paper and outlines directions for future research. Finally,  appendix~\ref{appendixA} provides supporting documentary figures for reference, in case the associated online profiles are modified or removed.

\section{The Phenomenon of Questionable Papers}
\label{ThePhenomenonofFakePapers}
Over the past couple of years, we have been conducting authentic research on the use of GenAI in ethical hacking. Our ultimate goal has been to enhance and automate the often manual and tedious penetration testing processes. We have published our findings and shared our advancements with the research community~\cite{TechReportUnAIInEH_HC_2024,2024_APracticalExaminationofManualExploitationPrivilegeEscalation_HC,2025_PenTest++_HC,STM24_UnleashingAIinEthicalHacking,ITASEC25_AdvancingEthicalHackingWthAIALinux-basedExperimentalStudy_2025}.

Interestingly, while following up on similar topics via Google Scholar alerts, we were recommended a group of articles that appeared —on the surface— to be highly relevant. However, upon close inspection, these papers struck us as being extremely thin in content and lacking technical depth or academic rigour. Based on their language, structure, and lack of empirical evidence, we strongly believe that these papers were entirely AI-generated and, in essence, fake.

To better understand the nature of the suspected potentially AI-generated papers, we performed a close textual and structural analysis of representative articles recommended to us via Google Scholar. At first glance, many of these papers appeared relevant to our ongoing work on AI and ethical hacking; they were well-formatted, readable, and addressed familiar themes. However, upon deeper examination, several red flags became apparent, as follows. 

\subsection{Key Concerns}
\begin{enumerate}
  \item \textbf{Many Published in a Single Snapshot in Time.} \\
  A striking observation was the publication date clustering. For instance, a significant number of these papers were released in the same month and year (e.g. February 2025~\cite{2025_IoTSecurityintheQuantumEra_Pomeroy,2025_EthicalHackingintheAgeofAIandIoT_RosethTom,2025_QuantumComputingandBlockchainSynergy_RosethTom,2025_AI-PoweredCyberSecurity_RosethTom,2025_MachineLearningforSOC_RosethTom,2025_LeveragingBlockchainForDecentralizedAndSecureSOCOperations_FakePaper,2025_QuantumComputingAndCybersecurity_SpundaFakePaper,2025_AdvancedEthicalHackingTechniques_Spunda,2025_SOCOptimizationThroughAI-PoweredAutomation_Spunda,2025_Blockchain-PoweredIoTSecurity_Pomeroy,2025_AIandQuantumComputing_Pomeroy,2025_EthicalHackingMeetsAI_Pomeroy,2025_StrengtheningSOCOperations_Pomeroy,hayat2025ethical_FakePaper,2025_TheRoleOfBlockchainSecuringIoTDevices_Spunda}), despite being attributed to different authors. This synchronicity implies the likelihood of batch generation and coordinated mass upload rather than independent scholarly efforts.

  \item \textbf{Strikingly Similar Content.} \\
  The papers displayed remarkably similar language, structure, and narrative flow. Phrases such as “AI-powered penetration testing enhances efficiency, accuracy, and scalability” were frequently reused across multiple documents, often with minimal variation. This degree of overlap is characteristic of AI-generated output or templated production rather than distinct academic voices.

  \item \textbf{Recycled and Dubious References.} \\
  The reference sections across these papers were nearly identical, containing overlapping citations with questionable relevance. Names like “Mohammed, A.” and “Żywiołek, J.” appeared recurrently, often in contexts that bore little relation to the main topic. Several citations mimicked DOI-style formatting but could not be verified via trusted academic databases, raising concerns about citation fabrication.

  \item \textbf{Absence of Methodology or Empirical Work.} \\
  The analysed papers contained no experimental sections, case studies, or data-driven evaluations. Their claims —such as “AI can automate ethical hacking tasks”— were presented without evidence or validation. The lack of methodology severely undermines their academic contribution.

  \item \textbf{No Engagement with Existing Literature.} \\
  Unlike authentic research papers, which usually position their contributions within the existing body of work, these questionable papers made no attempt to position their arguments within the broader academic discourse. There was no literature review or critical engagement with prior work, indicating a lack of scholarly depth and awareness.

  \item \textbf{Unverifiable Author Identities.} \\
  Basic searches for many of the listed authors yielded no academic profiles, institutional affiliations, or presence on reputable platforms like Google Scholar. The complete absence of online academic footprints for most of these names is highly suspicious and supports the hypothesis that the identities may be fabricated.
  
    \item \textbf{Predictable Filename Patterns.} \\
  We observed that many of these papers follow a noticeable pattern in their file naming conventions on ResearchGate, often using sequential and predictable numbering (e.g. 17, 18, 19, 20). This implies an automated or systematic upload process, further supporting the hypothesis of batch generation and artificial publication activity.

  \item \textbf{Non-Scientific Writing Style.} \\
  While the papers maintained surface-level readability, their tone resembled that of promotional blog posts rather than scientific discourse. They relied heavily on buzzwords like “AI revolutionising cybersecurity,” yet offered little technical specificity or conceptual clarity.
\end{enumerate}

\subsection{Final Verdict}

Based on these findings, we conclude that many of the papers in question are either AI-generated or strategically crafted with minimal effort for the sole purpose of inflating academic metrics. If subjected to standard peer-review, they would likely be rejected due to their lack of originality, insufficient methodology, and overall absence of scientific contribution.

Their presence on platforms like ResearchGate not only dilutes the quality of academic discourse but also highlights systemic vulnerabilities in how such platforms manage, curate, and verify scholarly content.

\section{Possible Motivations for Fabricated Paper Generation}
\label{PossibleMotivationsforFakePaperGeneration}
We now pose the question: Why are such papers being generated at such an alarming rate? Our hypothesis is that they exist solely to artificially inflate citation-based academic metrics, especially the H-index.

One example of where an academic's publication metrics appear to have benefitted from the upload of large numbers of questionable papers is as
follows.  We should point out that it may be entirely accidental that the individual concerned has benefitted in this way, since we cannot say for
sure why the papers have been created and uploaded; moreover, there are almost certainly many other cases we could have highlighted.  Our example
case is that of  Anwar Mohammed,  listed on Google Scholar as affiliated with Singhania University, Rajasthan, India. His profile\footnote{https://scholar.google.com/citations?hl=en\&user=gpUs8nkAAAAJ}  shows 
that a number of recently published articles of which he is an author have been cited in a number of apparently questionable preprints uploaded to ResearchGate. These preprints were automatically processed by Google Scholar, significantly changing its computation of metrics such as the H-index and i10-index.

This discovery strongly supports our hypothesis and prompted us to continue the investigation. Using Google Scholar’s reverse citation feature, we were able to uncover a chain of related questionable papers uploaded to ResearchGate. 
Our findings revealed a consistent pattern involving many ResearchGate members. Notably, we identified four illustrative cases —Spunda\footnote{https://www.researchgate.net/profile/Rudolf-Spunda-2/}, Pomeroy\footnote{https://www.researchgate.net/profile/Jennifer-Pomeroy-3/}, Gatlin\footnote{https://www.researchgate.net/profile/Kaiser-Gatlin/}, and Roseth\footnote{https://www.researchgate.net/profile/Tom-Roseth-3/}— each linked to a series of suspicious publications, many dated 2025~\cite{2025_IoTSecurityintheQuantumEra_Pomeroy,2025_EthicalHackingintheAgeofAIandIoT_RosethTom,2025_QuantumComputingandBlockchainSynergy_RosethTom,2025_AI-PoweredCyberSecurity_RosethTom,2025_MachineLearningforSOC_RosethTom,hayat2025ethical_FakePaper,2025_LeveragingBlockchainForDecentralizedAndSecureSOCOperations_FakePaper,2025_QuantumComputingAndCybersecurity_SpundaFakePaper,2025_AdvancedEthicalHackingTechniques_Spunda,2025_SOCOptimizationThroughAI-PoweredAutomation_Spunda,2025_TheRoleOfBlockchainSecuringIoTDevices_Spunda,2025_Blockchain-PoweredIoTSecurity_Pomeroy,2025_StrengtheningSOCOperations_Pomeroy,2025_AIandQuantumComputing_Pomeroy,2025_EthicalHackingMeetsAI_Pomeroy}, suggesting a deliberate and repeated engagement in questionable authorship practices. 
 Even more telling was the consistent authorship model:
\begin{itemize}
  \item \textbf{First author:} Typically an unfamiliar name, likely fictitious, often appearing only once or twice across the literature.
  \item \textbf{Second author:} A ResearchGate member with an active profile and prior publication history—frequently the actual uploader of the questionable paper.
\end{itemize}
To demonstrate this recurring pattern, we compiled a comparison table (see table ~\ref{ComparisonOfSuspectedFakeResearchGateProfiles}) with representative examples. Each case highlights how the second author role is used to anchor the fake paper with an ResearchGate-verified identity, while the likely fictitious first author provides plausible deniability. 
\begin{table}[ht]
\centering
\caption{Comparison of Suspected Questionable ResearchGate Profiles}
\label{ComparisonOfSuspectedFakeResearchGateProfiles}
\renewcommand{\arraystretch}{1.5}
\setlength{\tabcolsep}{5pt}
\begin{tabular}{||p{3.2cm}||p{3cm}||p{3cm}||p{3cm}||}
\hline\hline
\textbf{Feature} & \textbf{Rudolf Spunda} & \textbf{Jennifer Pomeroy} & \textbf{Kaiser Gatlin}   \\
\hline\hline

ResearchGate Profile & 
\url{https://www.researchgate.net/profile/Rudolf-Spunda-2/research} & 
\url{https://www.researchgate.net/profile/Jennifer-Pomeroy-3/research} & 
\url{https://www.researchgate.net/profile/Kaiser-Gatlin/research}  \\
\hline
Total Publications on ResearchGate & 10 & 10 & 10  \\
\hline
Solo-Authored Papers & 4 & 5 & 7 \\
\hline
Papers as 2nd Author & 6 & 5 & 3  \\
\hline
 Ephemeral 1st Authors & 
Fani Sani, Kuldeep Rahul, Abhishek Sharma, Muhammad Shamir, Yasir Nawaz, Haider Ali, Muhammad Nasir, Abrar Ahmed, Mukesh Kumar, Asia Rehman, Sadia Farooq, Halima Sinisa & 
Umair Zafer, Andre Russell, Ravi Bishnoi, Talaat Hayat, Nimra Fahad, Lennon Zahir & 
Ybk Jaiswal, K. L. Yadav, Shahbaz Ahmed \\
\hline
2nd Author Name (ResearchGate Member) & Rudolf Spunda & Jennifer Pomeroy & Kaiser Gatlin  \\
\hline
Profile Upload Activity & Uploaded by Rudolf & Uploaded by Jennifer & Uploaded by Kaiser   \\
\hline
Citation Count (on ResearchGate) & 1 & Not specified & 1  \\
\hline
Notes / Suspicious Behaviour & 
Uses ResearchGate to legitimise papers by collaborating with obscure authors & 
Repeats similar publication strategy across random names & 
Maintains consistent number of uploads, similar patterns, and solo-authorships  \\
\hline\hline
\end{tabular}
\end{table}

\section{Preliminary Experiment}
\label{Preliminary Experiment}

Given the gravity of the implications uncovered during our investigation, we designed a controlled experiment to determine whether the suspicious behaviours observed on ResearchGate could be deliberately reproduced and exploited using GenAI. Our aim was not only to test the platform’s safeguards but also to empirically demonstrate the mechanisms by which citation manipulation might occur in practice.

\subsection{Objectives}

The experiment was guided by the following objectives:

\begin{enumerate}
  \item To test whether a fake academic paper could be easily generated entirely by GenAI.
  \item To test whether a fake, AI-generated academic paper could be successfully uploaded to ResearchGate.
  \item To assess whether this paper, by citing other suspected questionable papers, could contribute to inflating citation metrics such as the H-index for their listed authors.
  \item To observe whether ResearchGate has any detection mechanisms or moderation workflows capable of identifying and removing inauthentic or AI-generated content.
\end{enumerate}

\subsection{Paper Generation Using GenAI}

We began by using GenAI tools (ChatGPT\footnote{https://chatgpt.com/} in particular) to generate a completely fictional academic paper titled \textit{``GenAI in Digital Forensics: Enhancing Timeline Reconstruction and Anomaly Attribution''}\footnote{\url{http://dx.doi.org/10.13140/RG.2.2.35588.23688}}. The content was generated with minimal human editing to preserve the authenticity of the experiment and to reflect what a typical misuse of GenAI might look like in a real-world setting. The output was fluent, well-structured, and mimicked the format of legitimate academic writing.

To ensure transparency, we embedded two disclaimers in the document: one on the title page; and another at the end. Both disclaimers clearly stated that the paper was generated using GenAI for academic experimentation and research purposes only, with no intent to deceive.

The generated paper lacked empirical evidence, literature review, or technical depth, intentionally mirroring the characteristics we had previously observed in suspected fake publications. The language was generalised, and the conclusions were deliberately vague, echoing the stylistic markers of questionable submissions we had reviewed.

\subsection{Submission Process and Citation Strategy}

Once the paper was prepared, we proceeded to upload it to ResearchGate. One key challenge  we encountered was that ResearchGate required at least one of the listed authors to be associated with an existing and verified ResearchGate profile. To bypass this requirement without linking the paper to our real identities, we introduced a third fictitious author, \textit{Naif Al-Sinani}, sharing a surname with the uploader. This manoeuvre satisfied the platform's verification checks.

Although the uploaded PDF listed only fictitious names as authors, ResearchGate automatically linked the document to the real profile used for submission, revealing a structural flaw in its authorship verification system.  At no point did the platform request institutional credentials or validate author identities against external academic databases—apart from an optional offer to invite the fictitious co-authors to join ResearchGate, which we simply ignored. 

To simulate real-world citation manipulation strategies, we embedded citations to at least three previously identified questionable papers, each authored or co-authored by ResearchGate users Spunda, Pomeroy, Gatlin, and Roseth. This was done deliberately to test whether our fake paper would successfully contribute to boosting their citation metrics.

\subsection{Initial Results and Observations}

At the time of writing, the fake paper\footnote{\url{http://dx.doi.org/10.13140/RG.2.2.35588.23688}} remains publicly accessible and has not been flagged, moderated, or removed by ResearchGate. Notably, within 24 hours of publication, we observed an increase in the citation counts of the questionable papers cited, suggesting that ResearchGate’s impact metrics are responsive to such fabricated references, thereby enabling H-index manipulation through a feedback loop.

\subsection{Limitations of the Study}

While the findings of our preliminary experiment offer valuable insights into the potential for citation manipulation on ResearchGate, several limitations must be acknowledged.

First, the experiment was intentionally limited in scope and designed as a single-case study. We uploaded only one fake paper, generated using a specific GenAI tool (ChatGPT), and assessed its effects over a short time frame. As such, the generalisability of our observations to broader contexts—such as across disciplines or different types of fake content—remains constrained.

Second, we did not actively attempt to scale or repeat the experiment with variations in author names, content quality, or citation strategies. A larger sample of fake papers with diverse formats and submission patterns would be necessary to draw stronger conclusions about the robustness of ResearchGate’s detection mechanisms.

Third, our experiment was conducted within the ethical bounds of academic research. For that reason, we refrained from citing our own prior work or identities to avoid metric inflation. This self-imposed constraint limits the scope of our conclusions about how self-citation dynamics may affect authorship metrics under malicious conditions.

Fourth, the effects we observed—such as the rise in citation counts—may be influenced by external variables (e.g., ResearchGate indexing policies, visibility algorithms) that we could not control or fully account for. Additionally, the increase in citation count observed in the cited papers may not necessarily reflect longer-term metric manipulation or sustained academic recognition.

Finally, our findings are specific to ResearchGate and may not directly apply to other platforms such as Academia.edu, arXiv, or SSRN, each of which may employ different content moderation and verification policies.

Despite these limitations, the study provides concrete, empirical evidence that ResearchGate's current infrastructure can be exploited to publish GenAI-generated content and manipulate citation-based metrics. Future work should address these limitations by extending the study longitudinally, incorporating multiple test cases, and collaborating with platform stakeholders for deeper analysis.

\subsection{Ethical Considerations}

To remain within acceptable ethical boundaries, we deliberately chose not to expand the experiment beyond a minimal intervention. No real identities, publications, or institutional affiliations were cited, and disclaimers were added for full transparency. Indeed, none of our real publications or identities were cited to avoid personal metric inflation.

Although the experiment was limited in scope, the results are sufficient to confirm that ResearchGate’s current platform design can be exploited to generate and circulate fake academic content with measurable impact on citation-based metrics. We plan to inform relevant stakeholders, including ResearchGate and Google Scholar, and will consider further controlled experiments in collaboration with academic institutions and ethical review boards.

\section{Implications and Recommendations}
\label{ImplicationsAndRecommendations}

The proliferation of fake academic papers poses a direct threat to the credibility of digital scholarly platforms and the academic community at large. Our findings reveal that individuals are exploiting the goodwill and open-access ethos of platforms like ResearchGate to artificially boost their academic profiles. By generating low-quality or AI-produced content and embedding self-referential citation loops, these actors are able to manipulate key academic metrics —especially the H-index— without contributing any real scholarly value.

Such practices erode the trust that underpins academic publishing. Platforms like ResearchGate, which were created to democratise access to research and foster global collaboration, are now being used as tools for metric-based self-promotion. This not only undermines the fairness of scholarly competition but also distorts the academic record, making it harder for genuine work to be recognised amidst a flood of fabricated content.

\subsection{Impact on Academic Metrics: H-index Gaming}

The central risk lies in the ability of fake papers to distort citation-based impact indicators. Since ResearchGate assigns DOIs to uploaded papers —and these DOIs are often indexed by third-party platforms such as Google Scholar— citations from these questionable documents contribute directly to the H-index and i10-index of the cited authors, regardless of the citing paper’s authenticity.

This phenomenon was clearly observed in the case of Anwar Mohammed, whose Google Scholar profile shows a disproportionate spike in citations originating from suspicious ResearchGate uploads, most dated 2024–2025. These documents follow consistent structural and stylistic patterns and frequently cite Mohammed’s earlier work, despite lacking any thematic relevance or scholarly depth.

Such gaming of citation metrics not only misrepresents the academic influence of certain individuals but also skews global rankings, hiring decisions, grant evaluations, and editorial opportunities, many of which still rely heavily on bibliometric indicators.

\subsection{Consequences for Academic Integrity and Content Quality}

The spread of fake publications has broader consequences beyond individual metric inflation. It compromises the overall quality of accessible research and confuses scholars —especially students and early-career researchers— who may struggle to differentiate legitimate contributions from AI-generated or fraudulent work. The appearance of scholarly legitimacy, conferred by DOIs, academic formatting, and citation counts, can mask a complete absence of methodological rigour or empirical contribution.

Moreover, the presence of unverified and unauthored content on scholarly platforms creates an uneven playing field, devalues peer-reviewed research, and undermines institutional trust in open repositories. If left unchecked, this trend could trigger a larger crisis of credibility for digital knowledge platforms.

\subsection{Ethical Considerations and Platform Responsibility}

Our investigation raises serious ethical and operational concerns:

\begin{itemize}
  \item Should platforms like ResearchGate be permitted to assign DOIs to unverified, unreviewed documents?
  \item Is it appropriate for citation metrics to be derived from documents that have not undergone any form of academic vetting?
  \item Do platforms have a duty to proactively prevent the use of GenAI for deceptive academic practices?
\end{itemize}

While our own experiment was conducted transparently and ethically —with disclaimers and no self-citations— the very fact that it was feasible demonstrates how easily such systems can be abused by bad actors. The onus now falls on platform providers to implement better safeguards, and on the academic community to hold such platforms accountable.

\subsection{Policy Recommendations}

In light of the findings presented in this study, we propose a series of platform-level and community-wide reforms:

\begin{itemize}
  \item \textbf{Stricter authorship verification}: Require institutional email addresses or ORCID verification for all listed authors before permitting uploads.
  \item \textbf{Content authentication systems}: Develop automated tools to flag GenAI-generated submissions using linguistic fingerprinting and structural pattern recognition.
  \item \textbf{Transparency labelling}: Introduce tags or visibility filters to identify papers marked as `experimental,' `non-peer-reviewed,' or `AI-generated,' and exclude them from impact metric calculations.
    \item \textbf{Exclude unfiltered preprints from citation metrics}: Google Scholar and other major sites providing information about research publications should stop using unfiltered preprints as a source for its computation of citation counts and metrics such as the H-index. Otherwise they become completely useless. 
  \item \textbf{Ethical review for mass uploads}: Require flagged high-volume uploaders to undergo manual review, especially in cases where sequential or templated publication behaviour is detected.
  \item \textbf{Community reporting mechanisms}: Enable and encourage users to flag suspicious documents or profiles for review, with transparent follow-up by platform moderators.
    \item \textbf{Preserving the integrity of DOIs}: Platforms should refrain from assigning DOIs to unvetted or automatically uploaded preprints. The indiscriminate use of DOIs in such cases dilutes their value and creates a false perception of scholarly legitimacy. 

\end{itemize}

\subsection{Call to Action}

The academic community must acknowledge that citation fraud, once confined to predatory journals and obscure repositories, is now infiltrating mainstream platforms. Combating this phenomenon will require a concerted effort from platform operators, institutions, funders, and individual researchers. Left unaddressed, the long-term consequences will be a gradual erosion of trust in digital scholarship, a distortion of merit-based academic recognition, and a dilution of legitimate scientific progress.

\section{Related Work}
\label{RelatedWork}

Recent work has demonstrated that citation counts on Google Scholar can be artificially inflated through paid services, highlighting a new and concerning form of metric  manipulation~\cite{2025_CitationManipulationThroughCitationMillsAndPreprintServers_HazemIbrahim}. This practice contaminates academic databases and misleads those who rely on citation metrics for scholarly evaluation.

The medical community has also explored the risks of AI-generated publications. M{\'a}jovsk{\'y} et al.~\cite{2023_ArtificialIntelligenceCanGenerateFraudulentButAuthenticLookingScientificMedicalArticlesPandoraBoxHasBeenOpened_Medical} showed how large language models (LLMs) like ChatGPT can be used to produce plausible but entirely fraudulent medical research articles. These papers were well-structured and readable, yet lacked any empirical grounding or peer-reviewed validation.

Earlier manifestations of this problem were seen with SCIgen, a program developed to auto-generate academic-looking computer science papers~\cite{2005_SCIGenAnAutomaticCSPaperGenerator_JeremyStribling}. Though initially satirical, SCIgen’s outputs were accepted at real conferences, providing early evidence that even minimal scrutiny could fail to detect synthetically generated nonsense.

These studies collectively underscore the systemic vulnerabilities of academic publishing, especially in the era of GenAI, and the urgent need for platforms, researchers, and policymakers to act against citation fraud and fake research dissemination.

This paper complements this prior work by providing a focused, experimental investigation into how AI-generated papers can be used to manipulate citation metrics on ResearchGate. We first identified clusters of questionable, AI-generated papers in a domain closely aligned with our own ongoing research on AI in ethical hacking. Notably, we observed that these papers disproportionately cited the work of a key beneficiary —Anwar Mohammed of Singhania University— whose citation metrics appear to benefit significantly from this citation pattern. Through a controlled upload and citation strategy, we highlight specific weaknesses in platform moderation and contribute empirical evidence to the broader discussion on academic integrity in the age of GenAI.

\section{Conclusions and Directions for Future Work}
\label{ConclusionandFutureWork}

\subsection{Conclusions and Findings}
This study investigated the presence and implications of questionable, possibly entirely AI-generated academic papers on widely used preprint repositories, such as ResearchGate, with a particular focus on their role in citation manipulation. Through detailed analysis and a controlled experiment, we confirmed that it is both technically and procedurally feasible to upload a synthetic paper generated using GenAI, link it to a legitimate ResearchGate profile, and use it to artificially boost the citation counts of other suspect publications. Our findings highlight structural weaknesses in platform moderation and reveal how these vulnerabilities can be exploited to manipulate bibliometric indicators such as the H-index and i10-index.

By examining patterns of suspicious publication behaviour, mapping citation loops, and engaging with ResearchGate’s content submission pipeline, we demonstrated how GenAI tools can be co-opted for metric-based academic fraud. Notably, our study exposes how seemingly legitimate citation activity—when powered by inauthentic documents—can distort scholarly metrics and rankings, ultimately undermining trust in academic evaluation systems.

While our study was limited in scope and designed with transparency and ethical safeguards, its implications are broader. The experiment not only reveals gaps in ResearchGate’s content validation mechanisms but also reflects a growing trend across scholarly platforms where ease of access and lack of verification are exploited for self-promotion. Our contribution adds empirical weight to the emerging literature on the risks posed by GenAI in academic publishing and calls for a collective reassessment of how digital scholarship is validated and quantified.

\subsection{Directions for Future Work}

Several avenues remain for further investigation. First, we plan to monitor the ongoing citation activity of the uploaded experimental paper, to assess the longer-term impact on the citation metrics of the documents it referenced. Second, we aim to extend our investigation to other widely used academic sharing platforms such as arXiv, Academia.edu, and SSRN, to evaluate whether similar vulnerabilities exist across ecosystems.

Additionally, we intend to engage in dialogue with platform providers like ResearchGate and indexing services such as Google Scholar to share our findings and advocate for the introduction of better safeguards. These may include stricter author verification procedures, metadata validation, and algorithmic detection of repetitive or AI-generated content.

On the institutional level, we plan to collaborate with academic stakeholders to help define ethical boundaries for the use of GenAI in research writing, and to contribute to the development of detection protocols for identifying fabricated academic content. In doing so, we hope to support the creation of more robust and trustworthy digital publishing environments.

Ultimately, this work aims to contribute to a larger effort of protecting the credibility of academic communication by exposing current weaknesses and motivating systemic change in how scholarly impact is measured and validated.

\section*{Disclaimer}
This study involved the creation and upload of a fake academic paper for experimental and ethical research purposes only. All actions were conducted with transparency and disclaimers, and no real individuals or institutions were misrepresented.

\section*{Acknowledgements}
Some portions of this manuscript were refined using GenAI tools (specifically, ChatGPT) to assist with language polishing and structural clarity. Following the use of such  tools, the authors thoroughly reviewed and edited the content as necessary and take full responsibility for the final publication. All core ideas, findings, experiments, and arguments were developed and authored by the named contributors.

\bibliographystyle{splncs04}
\bibliography{database}

\begin{thebibliography}{10}
\providecommand{\url}[1]{\texttt{#1}}
\providecommand{\urlprefix}{URL }
\providecommand{\doi}[1]{https://doi.org/#1}

\bibitem{2025_IoTSecurityintheQuantumEra_Pomeroy}
Ahmed, A., Pomeroy, J.: {IoT} security in the quantum era: Leveraging {AI} for
  predictive threat intelligence. ResearchGate (Feb 2025).
  \doi{10.13140/RG.2.2.32533.44002}

\bibitem{2025_QuantumComputingandBlockchainSynergy_RosethTom}
Ahmed, S., Roseth, T.: Quantum computing and blockchain synergy: A new paradigm
  for information security. ResearchGate (Feb 2025).
  \doi{10.13140/RG.2.2.22886.54083}

\bibitem{TechReportUnAIInEH_HC_2024}
Al-Sinani, H., Mitchell, C.: Unleashing {AI} in ethical hacking: A preliminary
  experimental study. Technical report, Royal Holloway, University of London
  (2024),
  \url{https://pure.royalholloway.ac.uk/files/58692091/TechReport_UnleashingAIinEthicalHacking.pdf}

\bibitem{2024_APracticalExaminationofManualExploitationPrivilegeEscalation_HC}
Al{-}Sinani, H.S., Mitchell, C.J.: {AI}-augmented ethical hacking: {A}
  practical examination of manual exploitation and privilege escalation in
  {Linux} environments. CoRR  \textbf{abs/2411.17539} (2024),
  \url{https://doi.org/10.48550/arXiv.2411.17539}

\bibitem{2025_PenTest++_HC}
Al{-}Sinani, H.S., Mitchell, C.J.: Pentest++: Elevating ethical hacking with
  {AI} and automation. CoRR  \textbf{abs/2502.09484} (2025),
  \url{https://doi.org/10.48550/arXiv.2502.09484}

\bibitem{STM24_UnleashingAIinEthicalHacking}
Al{-}Sinani, H.S., Mitchell, C.J., Sahli, N., Al{-}Siyabi, M.: Unleashing {AI}
  in ethical hacking. In: Martinelli, F., Rios, R. (eds.) Security and Trust
  Management --- 20th International Workshop, {STM'24}, Bydgoszcz, Poland,
  September 19--20, 2024, Proceedings. Lecture Notes in Computer Science, vol.
  15235, pp. 140--151. Springer (2024),
  \url{https://doi.org/10.1007/978-3-031-76371-7\_10}

\bibitem{ITASEC25_AdvancingEthicalHackingWthAIALinux-basedExperimentalStudy_2025}
Al{-}Sinani, H.S., Sahli, N., Mitchell, C.J., Al{-}Siyabi, M.: Advancing
  ethical hacking with {AI: A Linux}-based experimental study. In: Costa, G.,
  Montanari, R., Carminati, M., Sciarretta, G. (eds.) Joint National Conference
  on Cybersecurity (ITASEC \& SERICS 2025), February 03--08, 2025, Bologna,
  Italy, Proceedings. p. to appear (2025),
  \url{https://www.chrismitchell.net/Papers/aehwaa2.pdf}

\bibitem{2025_AIandQuantumComputing_Pomeroy}
Bishnoi, R., Pomeroy, J.: {AI} and quantum computing: Transforming information
  security protocols for the future. ResearchGate (Feb 2025).
  \doi{10.13140/RG.2.2.19111.66722}

\bibitem{2025_LeveragingBlockchainForDecentralizedAndSecureSOCOperations_FakePaper}
Fahad, N., Gatlin, K.: Leveraging blockchain for decentralized and secure {SOC}
  operations. ResearchGate (Feb 2025). \doi{10.13140/RG.2.2.32376.15360}

\bibitem{hayat2025ethical_FakePaper}
Hayat, T., Gatlin, K.: {AI}-powered ethical hacking: Rethinking cyber security
  penetration testing. ResearchGate (Feb 2025).
  \doi{10.13140/RG.2.2.18954.38080},
  \url{https://www.researchgate.net/publication/390128254_AI-Powered_Ethical_Hacking_Rethinking_Cyber_Security_Penetration_Testing}

\bibitem{2025_CitationManipulationThroughCitationMillsAndPreprintServers_HazemIbrahim}
Ibrahim, H., Liu, F., Zaki, Y., Rahwan, T.: Citation manipulation through
  citation mills and pre-print servers. Scientific Reports  \textbf{15}(1),
  ~5480 (Feb 2025), \url{https://www.nature.com/articles/s41598-025-88709-7}

\bibitem{2025_MachineLearningforSOC_RosethTom}
Jaiswal, Y., Roseth, T.: Machine learning for {SOC} operations: Automating
  threat detection and response. ResearchGate (Feb 2025).
  \doi{10.13140/RG.2.2.12820.21127}

\bibitem{2025_EthicalHackingintheAgeofAIandIoT_RosethTom}
Kumar, M., Roseth, T.: Ethical hacking in the age of {AI} and {IoT}: Proactive
  cyber defense strategies. ResearchGate (Feb 2025).
  \doi{10.13140/RG.2.2.26241.98400}

\bibitem{2023_ArtificialIntelligenceCanGenerateFraudulentButAuthenticLookingScientificMedicalArticlesPandoraBoxHasBeenOpened_Medical}
M{\'a}jovsk{\'y}, M., {\v{C}}ern{\'y}, M., Kasal, M., Komarc, M., Netuka, D.:
  Artificial intelligence can generate fraudulent but authentic-looking
  scientific medical articles: Pandora's box has been opened. J Med Internet
  Res  \textbf{25},  e46924 (May 2023). \doi{10.2196/46924},
  \url{https://www.jmir.org/2023/1/e46924}

\bibitem{2025_EthicalHackingMeetsAI_Pomeroy}
Nasir, M., Pomeroy, J.: Ethical hacking meets {AI}: Revolutionizing
  vulnerability assessments and penetration testing. ResearchGate (Feb 2025).
  \doi{10.13140/RG.2.2.25822.55368}

\bibitem{2025_AdvancedEthicalHackingTechniques_Spunda}
Rahul, K., Spunda, R.: Advanced ethical hacking techniques using {AI} and
  predictive modeling. ResearchGate (Feb 2025).
  \doi{10.13140/RG.2.2.20160.24326}

\bibitem{2025_StrengtheningSOCOperations_Pomeroy}
Russell, A., Pomeroy, J.: Strengthening {SOC} operations with blockchain for
  tamper-proof incident management. ResearchGate (Feb 2025).
  \doi{10.13140/RG.2.2.34211.16161}

\bibitem{2025_QuantumComputingAndCybersecurity_SpundaFakePaper}
Sani, F., Spunda, R.: Quantum computing and cybersecurity: Preparing for a
  post- quantum world. ResearchGate (Feb 2025).
  \doi{10.13140/RG.2.2.14288.21767}

\bibitem{2025_TheRoleOfBlockchainSecuringIoTDevices_Spunda}
Shamir, M., Spunda, R.: The role of blockchain in securing {IoT} devices and
  critical infrastructure. ResearchGate (Feb 2025).
  \doi{10.13140/RG.2.2.35259.73769}

\bibitem{2025_SOCOptimizationThroughAI-PoweredAutomation_Spunda}
Sharma, A., Spunda, R.: {SOC} optimization through {AI}-powered automation and
  blockchain integration. ResearchGate (Feb 2025).
  \doi{10.13140/RG.2.2.27709.99049}

\bibitem{2005_SCIGenAnAutomaticCSPaperGenerator_JeremyStribling}
Stribling, J., Krohn, M., Aguayo, D.: Scigen – an automatic cs paper
  generator (2005), \url{https://pdos.csail.mit.edu/archive/scigen/}

\bibitem{2025_AI-PoweredCyberSecurity_RosethTom}
Yadav, K., Roseth, T.: {AI}-powered cyber security: Enhancing {SOC} operations
  with machine learning and blockchain. ResearchGate (Feb 2025).
  \doi{10.13140/RG.2.2.10723.05927}

\bibitem{2025_Blockchain-PoweredIoTSecurity_Pomeroy}
Zafer, U., Pomeroy, J.: Blockchain-powered {IoT} security: Ensuring data
  integrity and device trustworthiness. ResearchGate (Feb 2025).
  \doi{10.13140/RG.2.2.15756.22404}

\end{thebibliography}

\appendix
\label{appendixA}
\section{Appendix: Supporting Screenshots}
\label{appendixA}
\addcontentsline{toc}{section}{Appendix: Supporting Screenshots}

\begin{figure}[ht]
\centering
\includegraphics[width=0.8\textwidth]{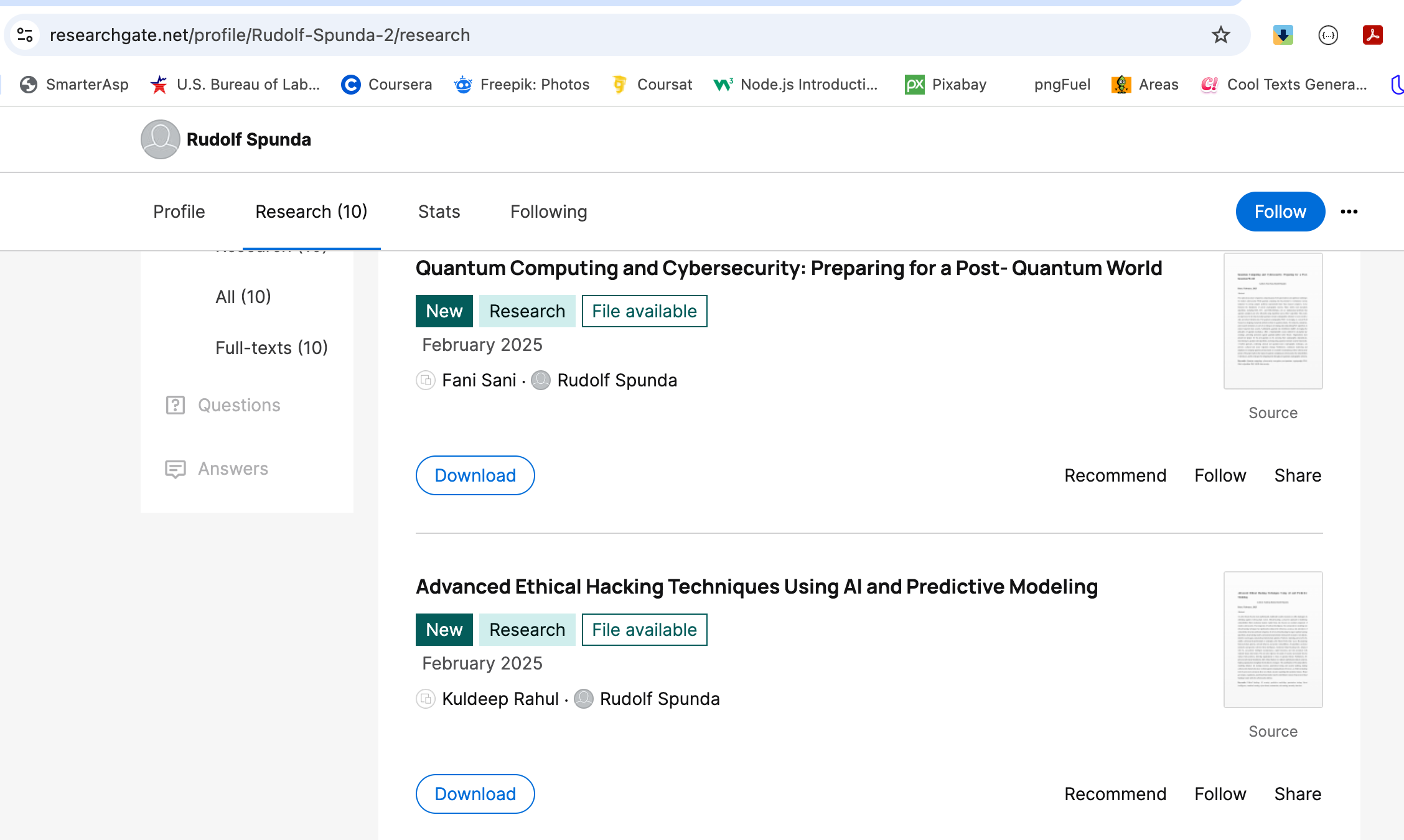}
\caption{Rudolf Spunda’s ResearchGate profile.}
\label{fig:spunda-rg}
\end{figure}

\begin{figure}[ht]
\centering
\includegraphics[width=0.8\textwidth]{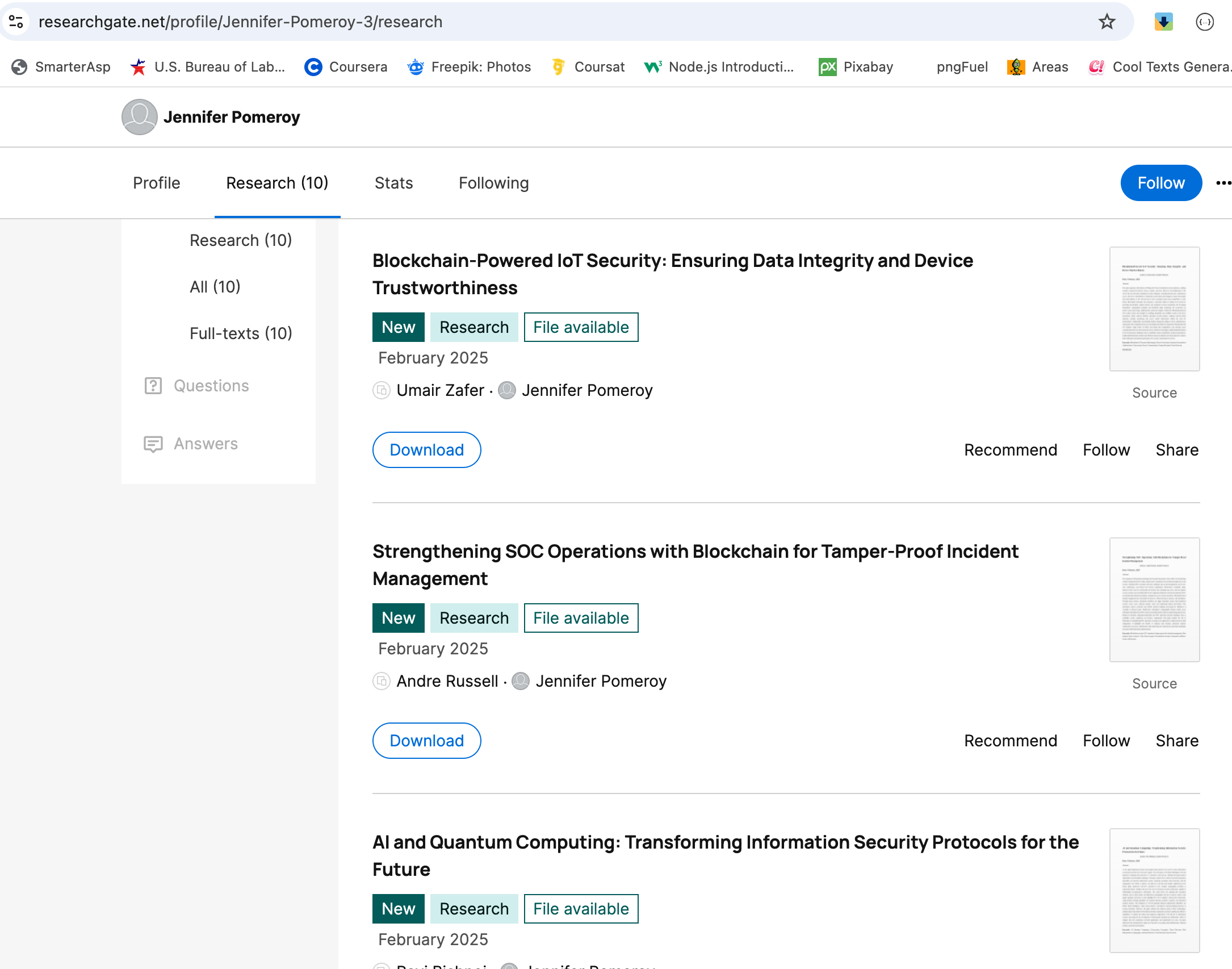}
\caption{Jennifer Pomeroy’s ResearchGate profile.}
\label{fig:pomeroy-rg}
\end{figure}

\begin{figure}[ht]
\centering
\includegraphics[width=0.8\textwidth]{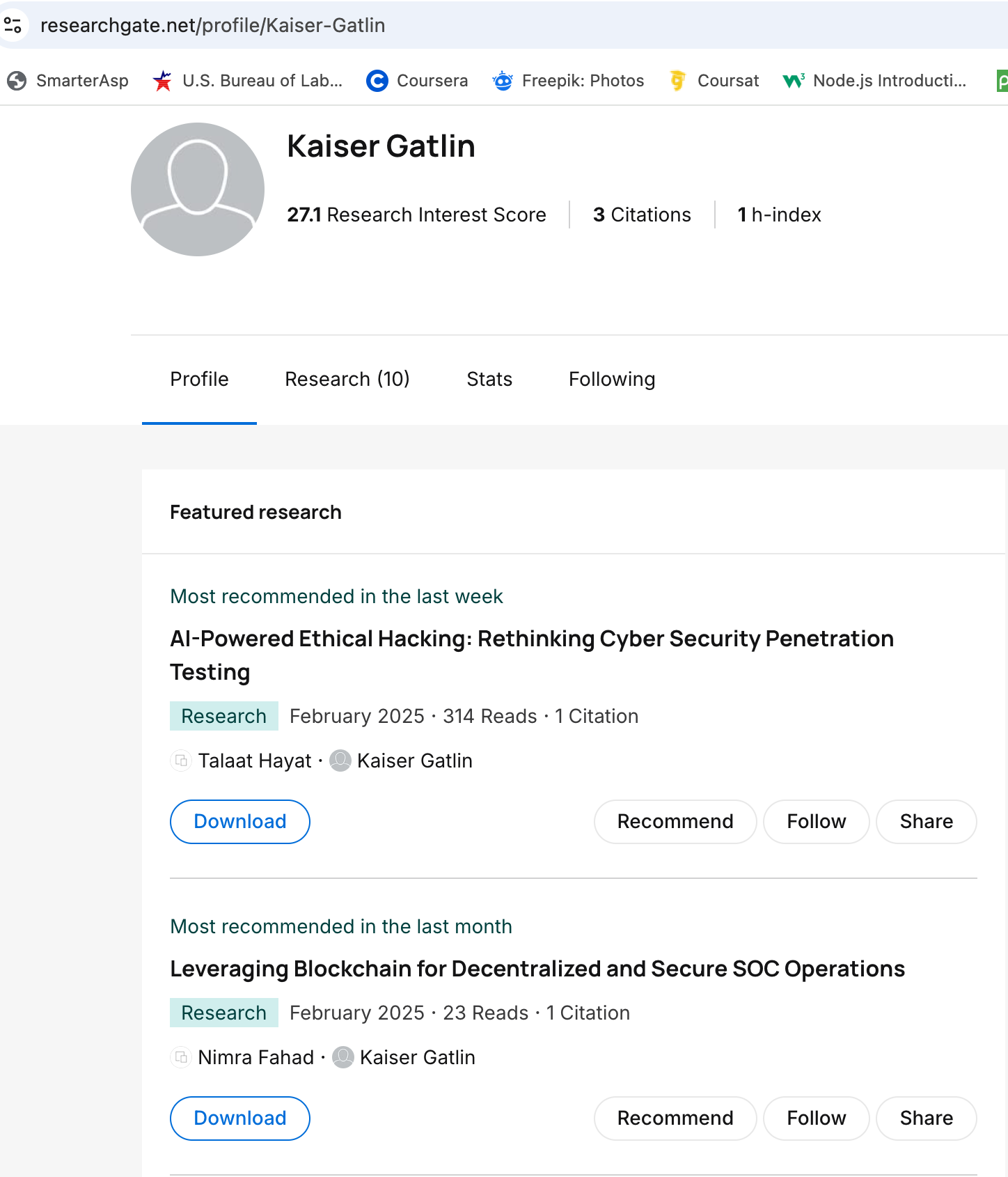}
\caption{Kaiser Gatlin’s ResearchGate profile.}
\label{fig:gatlin-rg}
\end{figure}

\begin{figure}[ht]
\centering
\includegraphics[width=0.8\textwidth]{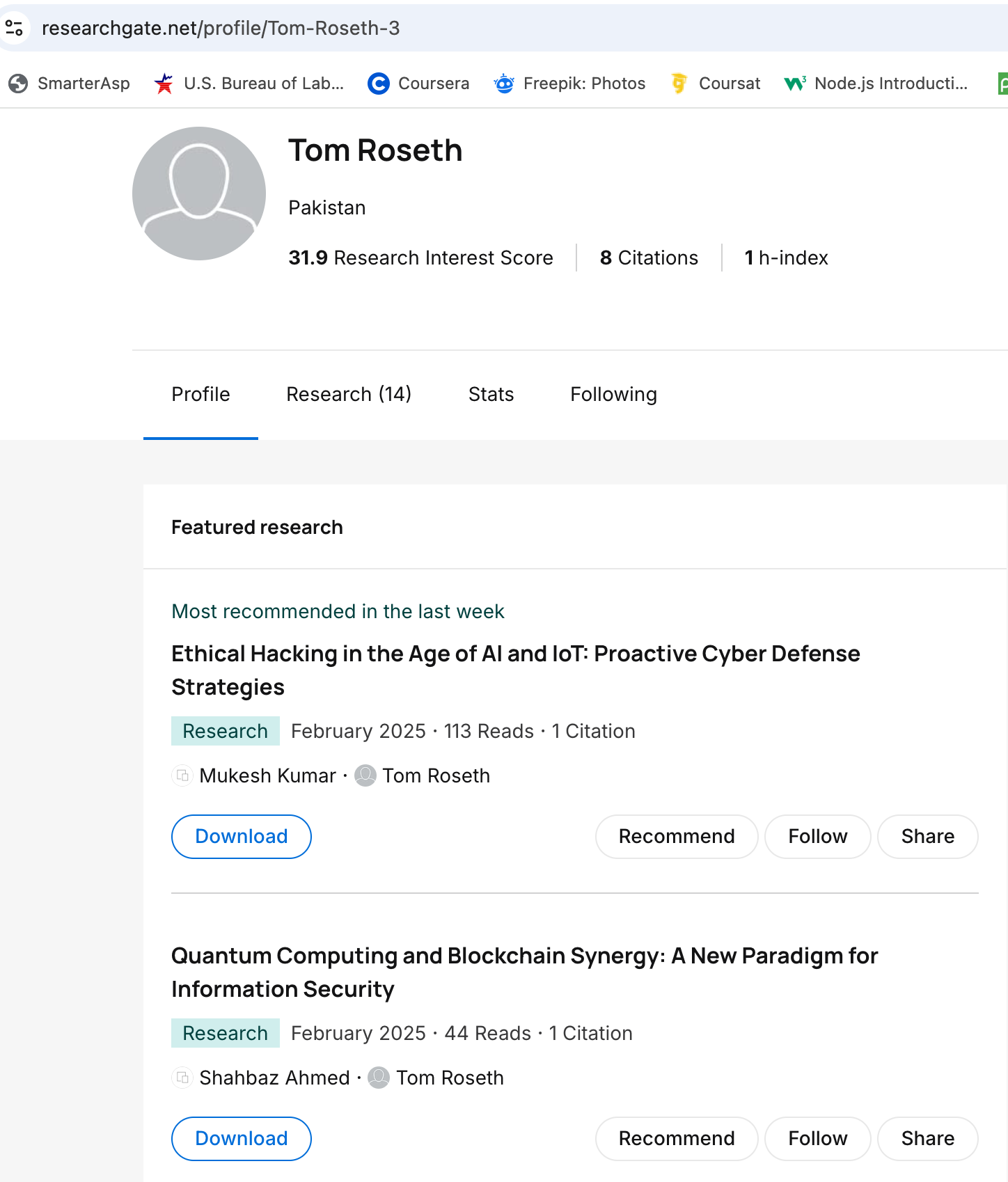}
\caption{Tom Roseth’s ResearchGate profile.}
\label{fig:roseth-rg}
\end{figure}

\begin{figure}[ht]
\centering
\includegraphics[width=0.8\textwidth]{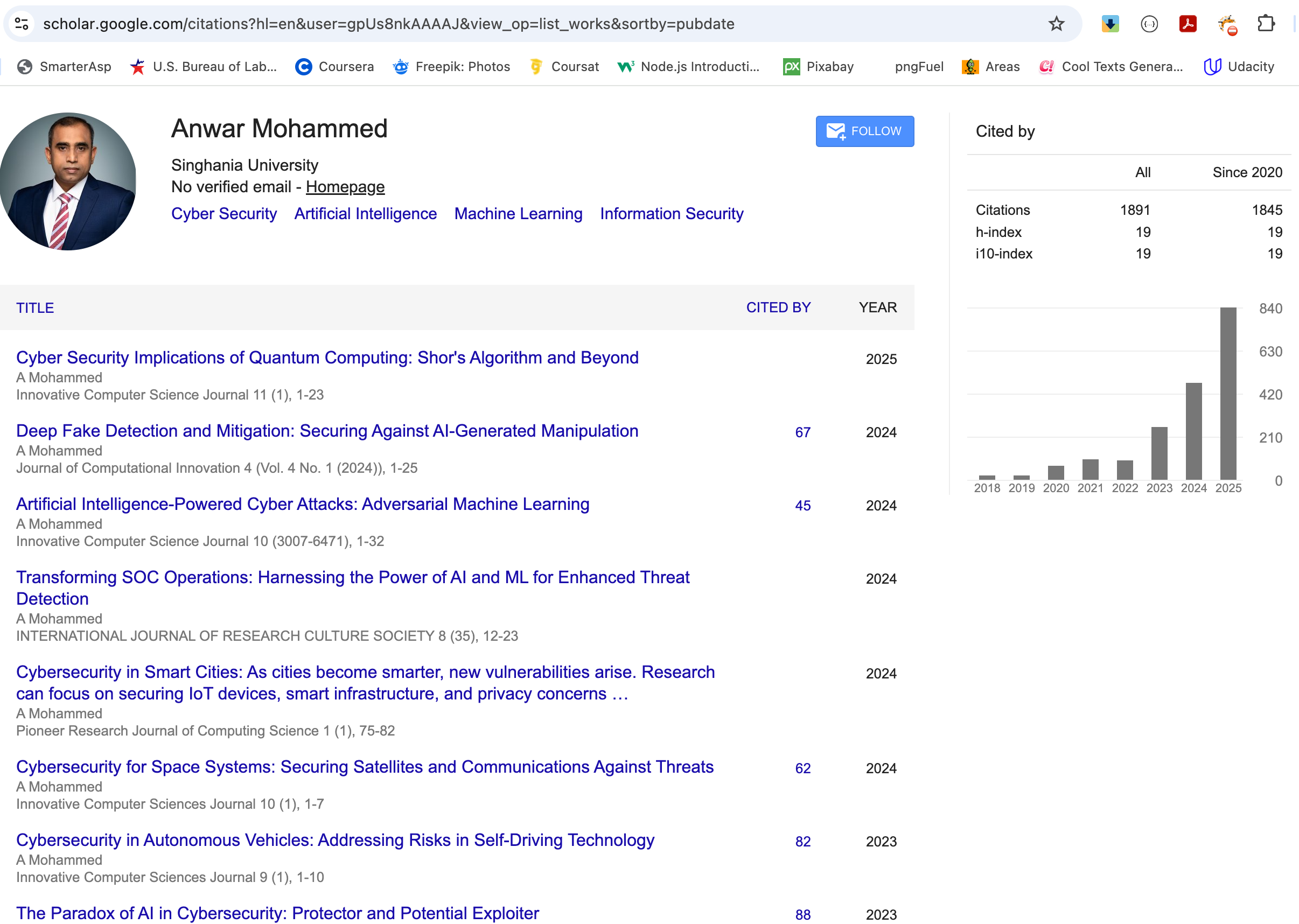}
\caption{Anwar Mohammed’s Google Scholar profile.}
\label{fig:anwar-scholar}
\end{figure}

\begin{figure}[ht]
\centering
\includegraphics[width=0.8\textwidth]{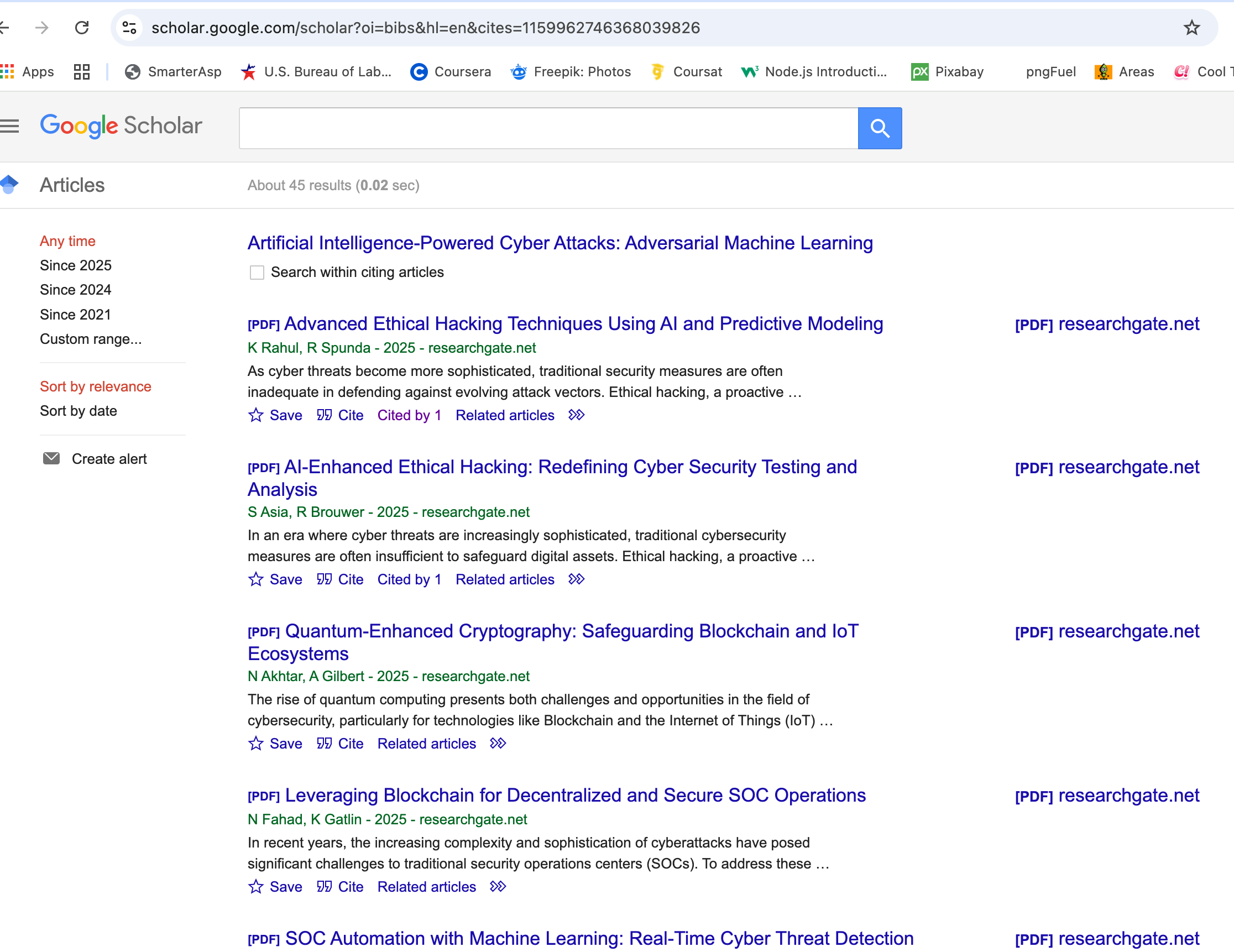}
\caption{Reverse citation trail from Google Scholar.}
\label{fig:reverse-cite}
\end{figure}

\end{document}